# Chemical sensing with atomically-thin metals templated by a two-dimensional insulator

Kyung Ho Kim[1], Hans He[1], Marius Rodner[2], Rositsa Yakimova[2], Karin Larsson[3], Marten Piantek[4,] David Serrate[5,6], Alexei Zakharov[7], Sergey Kubatkin[1], Jens Eriksson[2], Samuel Lara-Avila[1,8*]

[1] Department of Microtechnology and Nanoscience, Chalmers University of Technology, SE-412 96, Gothenburg, Sweden

[2] Department of Physics, Chemistry and Biology, Linkoping University, S-581 83, Linköping, Sweden.

[3] Department Chemistry-Angstrom Laboratory, University of Uppsala, Uppsala, Sweden

[4] Instituto de Nanociencia de Aragón and Laboratorio de Microscopías Avanzadas, Universidad de Zaragoza, 50018 Zaragoza, Spain.

[5] Instituto de Ciencia de Materiales de Aragón, CSIC-University of Zaragoza, 50009, Zaragoza, Spain

[6] Departamento de Física de la Materia Condensada, University of Zaragoza, 50009, Zaragoza, Spain

[7] MAX IV Laboratory and Lund University, 221 00 Lund, Sweden.

[8] National Physical Laboratory, Hampton Road, Teddington, TW11 0LW, UK

**Boosting the sensitivity of solid-state gas sensors by incorporating nanostructured materials as the active sensing element can be complicated by interfacial effects[1–4]. Interfaces at nanoparticles, grains, or contacts may result in non-linear current-voltage response, high electrical resistance, and ultimately, electric noise that limits the sensor read-out[5–8]. Here we report the possibility to prepare nominally one atom thin, electrically continuous metals, by straightforward physical vapor deposition on the carbon zero-layer grown epitaxially on silicon carbide. With platinum as the metal, its electrical conductivity is strongly modulated when interacting with chemical analytes, due to charges being transferred to/from Pt. This, together with the scalability of the material, allows us to microfabricate chemiresistor devices for electrical read-out of chemical species with sub part-per-billion detection limits. The two-dimensional system formed by atomically-thin metals open up a route for resilient and high sensitivity chemical detection, and could be the path for designing new heterogeneous catalysts with superior activity and selectivity.**

Chemiresistors represent a major class of commercial solid state gas detectors due to ease of fabrication, simple operation, low cost and possibility of miniaturization[2,4,8]. In these transducers, physicochemical interactions of the sensing material with gases can be simply monitored by a two-terminal resistance read-out[9,10]. With the advent of graphene, its use as chemical sensing element has led to outstanding sensitivities[3]. But while graphene is an exquisite charge sensor, it is a rather chemically inert material. Graphene's ability to interact with chemical species largely depends on its surface being functionalized with a sensitizing layer, such as metal/oxide nanoparticles[11–13], or even polymeric residues from the microfabrication process[3,14]. Yet, chemical bonds to graphene that might occur during functionalization –or operation at high temperature in the functionalized material– can disrupt its transduction capabilities. In atomically-thin crystals, covalent interaction with the crystal lattice can profoundly modify the band structure of the material, altering its electronic properties and with this, its charge sensing capabilities. This fragility towards chemical interactions of graphene, and two-dimensional (2D) crystals in general, might compromise taking full advantage of such prospective materials in emerging chemical sensing applications[12,15,16].

The chemical-to-electrical transduction with atomically-thin platinum that we present here is based on inherent interactions of transition metals with chemical analytes, and the fact that we are capable of preparing Pt in ultra-thin form, such that the resistance of the bulk-less metal is strongly dominated by surface effects. Thus, the metal acts simultaneously as the sensing and charge transducer layer. To prepare atomically-thin Pt layers we have used as the substrate the so-called carbon zero-layer, grown epitaxially on SiC (Fig. 1a). In contrast to other substrates[17–19], the insulating carbon zero-layer is not only key to enable the two-dimensional growth mode of Pt, but also to electrically probe the onset of conductivity of the surface at early stages of Pt deposition.

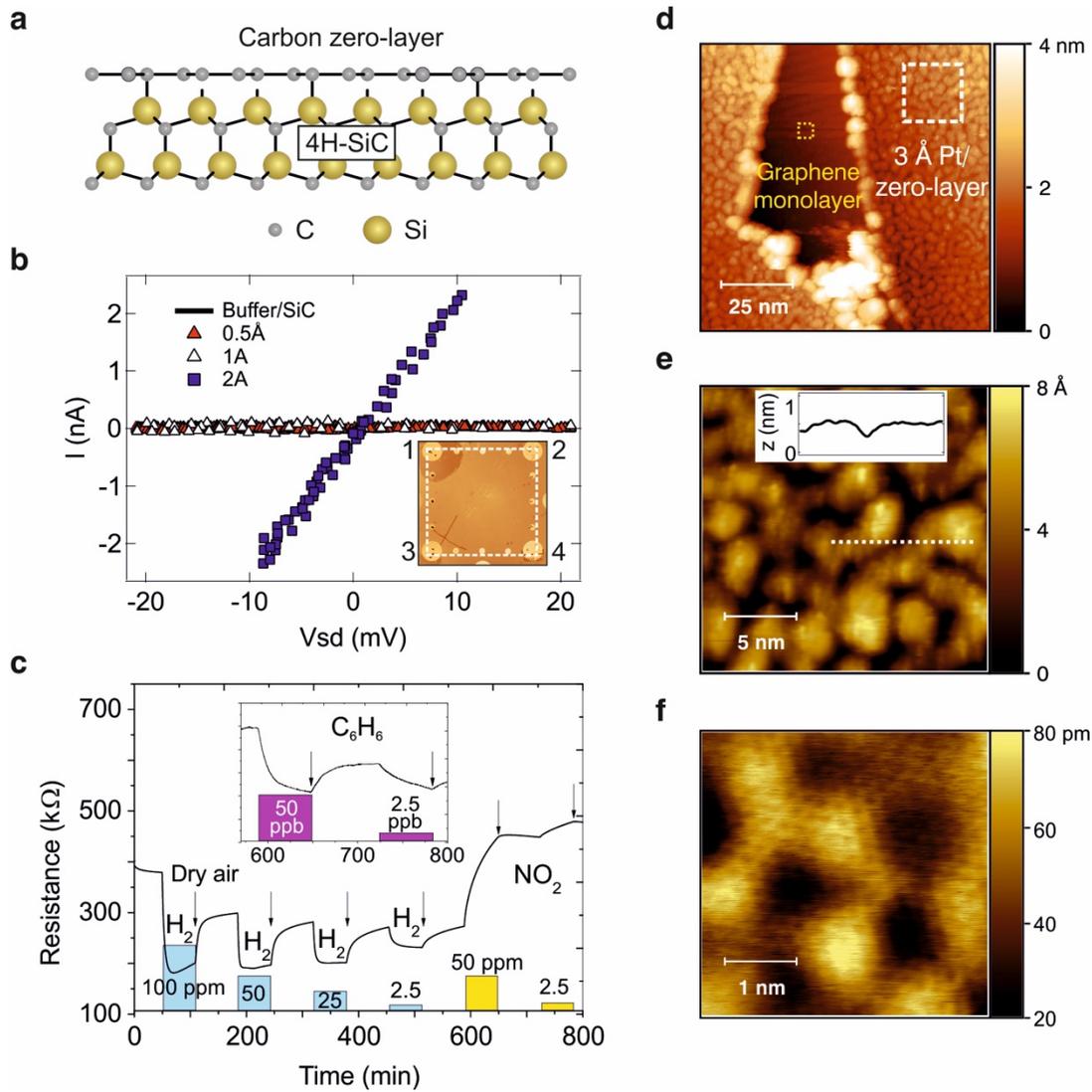

**Fig. 1. Atomically-thin Pt as gas sensor. a,** Schematic of the carbon zero-layer, epitaxially grown on the silicon terminated face (0001) of 4H-SiC. **b,** Current-voltage characteristic measured in-situ during Pt evaporation on a substrate at room temperature and nominal deposition rate r = 1 Å/min (background pressure P = 5 x 10$^{-7}$ mbar). Inset: optical micrograph of the 7x7 mm$^2$ chip used for in-situ resistance measurements; the resistance is measured between pre-deposited Pd contacts (circles) 1-2 and 3-4. Pt is deposited through a shadow mask with geometry indicated by the dashed white square. **c,** Sensing response of a 7 x 7 mm$^2$ substrate under exposure to $H_2$, $NO_2$, and $C_6H_6$ (inset) in a background of dry air. Arrows indicate the time at which the gas test chamber is purged with dry air. **d,** Constant current topography obtained by scanning tunneling microscopy (STM), giving an overview on a 100 nm x 100 nm area ($V_{STM}$ =1 V and $I_{STM}$=60 pA) of the Pt covered surface in the neighborhood of a graphene monolayer inclusion. **e,** STM scan (20 nm x 20 nm, $V_{STM}$ =-200 mV and $I_{STM}$= 200 pA) taken over a terrace (dashed white rectangle in (d)) showing the detailed topography of the Pt film. Inset: line profile indicated by the dotted line. **f,** STM scan ($V_{STM}$ =200 mV and $I_{STM}$= 200 pA) on the graphene monolayer inclusion showing that the graphene surface remains atomically clean after Pt deposition. The characteristic Moiré pattern of epitaxial graphene on SiC is evident in this atomically resolved scan.

Fig. 1b shows the current-voltage characteristic measured in-situ, as Pt deposition takes place at a slow rate (r = 0.017 Å/s), on a 7 x 7 mm$^2$ substrate held at room temperature. The initial high resistance measured on the zero-layer (R >> 10 GOhm) turns abruptly to an *ohmic*, linear current-voltage response after deposition of merely 2Å of Pt (for the complete R vs thickness see Supplementary Figure 1). The electrical resistance of Pt is particularly sensitive to the environment when the nominally deposited Pt thickness is t ≲ 1 nm. For instance, Fig. 1c shows the resistance changes of a nominally 3 Å thick Pt layer during exposure to standard test gases in the part-per-million range for strongly reducing ($H_2$) and strongly oxidizing ($NO_2$) gases. The inset of Fig. 1c also shows that electrical detection of chemical species with atomically thin Pt is readily possible in the part-per-billion range, using benzene as example (for formaldehyde see Supplementary Figure 2).

We investigated the surface of the Pt chemical transducer layer by combining scanning tunneling microscopy (STM) and electron transport measurements. Fig. 1d shows a 100 x 100 nm$^2$ STM image of the substrate surface after deposition of 3 Å Pt. This particular STM scan shows, in addition to the Pt-decorated zero-layer, a graphene monolayer inclusion which is a sub-product of the carbon zero-layer growth (for a larger STM scan see Supplementary Figure 3). On the zero-layer area, we observe that Pt forms a coalescing network of islands, about 3 Å high and predominantly 2D in character (Fig. 1e and inset therein). In sharp contrast, the surface of the graphene inclusion remains atomically clean after Pt deposition (Fig. 1f), revealing the stronger adhesion of Pt to the carbon zero-layer compared to that on graphene. Electrical transport measurements on Hall bar devices made with atomically-thin Pt, shown in Fig 2a, complement the scanning probe studies. First, the *metallic* resistance (dR/dT > 0) down to T ≈ 20 K, shown in Fig. 2b, implies that the Pt clusters observed by STM form a truly electrically interconnected network. Second, measurements in magnetic field (Fig. 2c), revealed that atomically-thin Pt is a hole-type conductor with a 2D hole density in the range $p_0$ =1-3x10$^{14}$ cm$^{-2}$, indicating carrier-asymmetric scattering at metal boundaries on this network[20,21,22].

Moreover, electrical characterization allows us to find the optimal conditions for the deposition, i.e. those yielding surfaces with the lowest resistance for a given Pt thickness. We found that a substrate temperature in the range 180 - 200 °C, and a much faster deposition rate (r = 1 Å/s) reproducibly yield resistivity in the range $\rho \approx 1 - 3$ kΩ/square for t = 3 Å ($\rho \lesssim 1$ kΩ/square for t = 4 Å) and the Pt morphology shown in Fig. 1e. The influence of the deposition parameters on the electrical resistance of atomically-thin Pt highlights the role of kinetic effects on the early stages of Pt growth on the zero-layer.

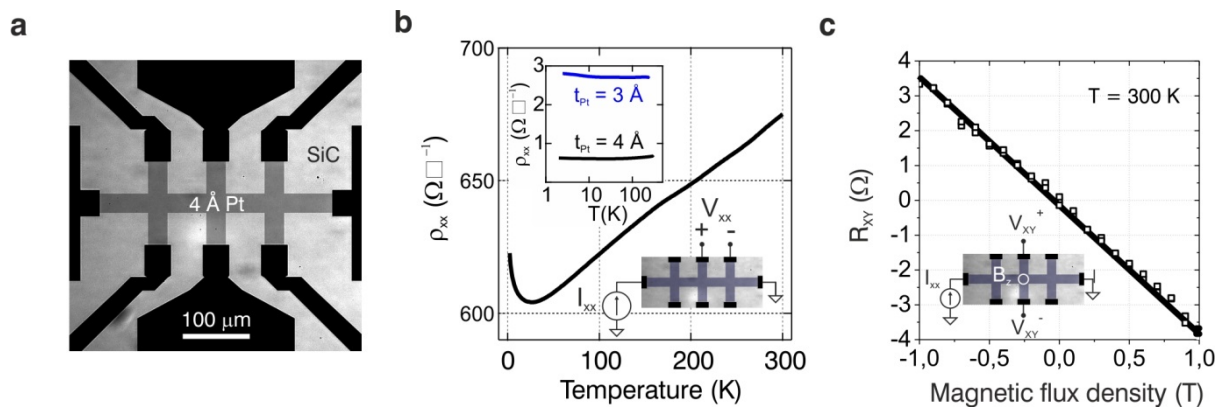

**Fig. 2. Electrical characterization of atomically-thin Pt deposited on the carbon zero-layer. a,** Optical micrograph in light-transmission mode of a Hall bar device (L = 180 μm x W = 30 μm) patterned on a nominally 4Å thick Pt film. **b,** Temperature dependence of the longitudinal resistivity $\rho_{xx} = \frac{R_{XX}}{(L/W)} = \left(\frac{W}{L}\right)\left(\frac{V_{XX}}{I_{XX}}\right)$ measured on the device shown in (a). The device resistance decreases down to T = 20 K. The increase of resistance for T < 20 K is due to quantum corrections to the classical resistance[36]. Inset: comparison of temperate dependence of resistivity for a 3Å and 4Å thick Pt-film **c,** Hall resistance $R_{XY} = V_{XY}/I_{XX}$ as a function of magnetic field $B$ reveals a negative Hall coefficient $R_H = dR_{XY}/dB$=-3.75 Ω/T, signaling a hole-type conductivity in Pt with a carrier density $p = 1/(eR_H) = 1.77 \times 10^{14}$ holes.cm$^{-2}$ for this device.

The carbon zero-layer is the crucial ingredient to ensure the early onset of electrical conductivity of Pt at low coverage (see control experiments on other substrates in Supplementary Figure 4). The carbon zero-layer is a highly ordered surface reconstruction of SiC resulting from thermal annealing at high temperature. If prepared properly, the zero-layer is truly a 2D crystal, graphene-like, in which about 30% of the carbon atoms are chemically bound to the SiC substrate. The partial chemical bonding to SiC prevents the development of linear π-bands in the zero-layer, rendering it electrically insulating[23–25]. We have studied the crystalline structure of our carbon zero-layer by low energy electron microscopy (LEEM), on substrates that led to low Pt sheet resistances (~1 kΩ/square). For LEEM studies, Pt is deposited only on half of the chip through a shadow mask so as to allow the comparison between the Pt-free and the Pt-covered zero-layer on the same substrate. The LEEM analysis shows that the Pt-free, pristine zero-layer displays a clear contrast between two types of domains, marked with yellow and blue circles in Fig.3a. Micro-low energy electron diffraction (µ-LEED), shown in Figures 3b,c, revealed that this contrast arises from the presence of two different rotated crystal domains corresponding each to the 6√3 x 6√3 carbon-rich phase of the zero-layer[24] (for Pt on the Si-rich SiC substrate see Supplementary Figure 5). In contrast, the Pt-covered area remains dark in LEEM across domains on the zero-layer, independently of electron energy, due to the lack of the long-range order of the Pt layer. This lack of range order in Pt, observed from LEEM, is consistent with our STM studies (Fig. 1d-e).

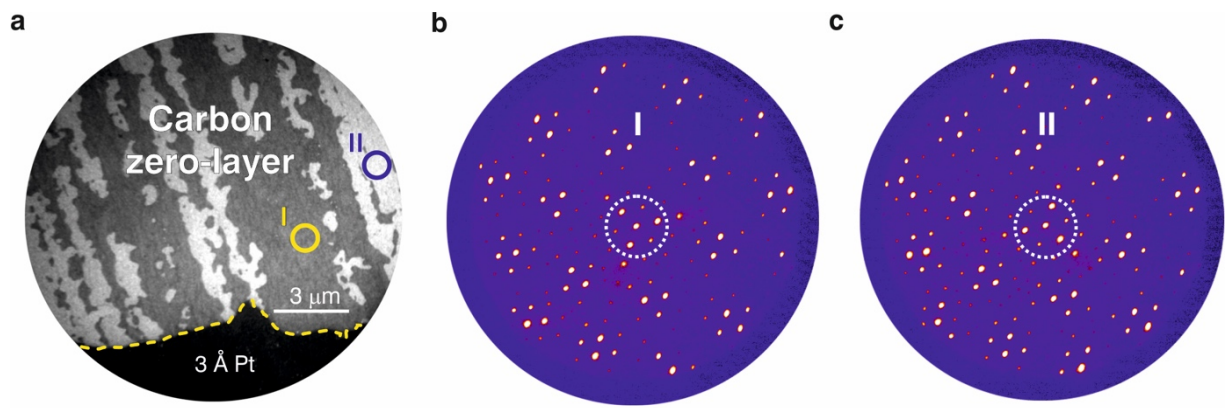

**Fig. 3. Surface characterization of atomically-thin platinum deposited on the carbon zero-layer. a,** Bright-field low energy electron microscopy (LEEM) image of the substrate taken at the boundary (dashed line) between the bare carbon zero-layer and the Pt-covered zero-layer. The Pt-covered surface (bottom part of the image) appears dark at all electron energies due to the lack of long-range ordering of the Pt layer. The contrast observed on the zero-layer (top part of the image), e.g. between the areas marked with yellow and blue circles in region I and II, show the presence of different zero-layer domains. **b, c,** Micro-low energy electron diffraction (LEED) images (E = 48 eV, sampling area 500 nm) collected on the zero-layer from regions I (yellow) and II (blue) in **a**, respectively, reveal that the LEEM contrast arises from the presence of two different rotated crystal domains (highlighted by the dashed white circle).

Our DFT calculations provide an insight on the origin of the quasi-2D growth mode of Pt on the zero-layer that allows us to achieve an atomically-thin metallic transducer. While kinetic constrains appear to determine the onset of electrical conductivity at low Pt coverage, the likelihood of a metal to wet the surface is ultimately given by thermodynamic considerations. At thermodynamic equilibrium, wetting is expected if the metal-substrate interfacial energy is lower than the difference between the surface free energy of the clean substrate and that of a clean metal[26]. With experimental surface free energy data not available for the zero-layer, we resorted to compare the theoretical adhesion energies of Pt atoms on free-standing graphene (G), the 6√3 x 6√3 carbon-rich zero-layer on SiC (ZL/SiC), and graphene/ZL/SiC (G/SiC) substrates. We define the adhesion energies as $\Delta E_{adhesion} = E_{surface-Pt} - E_{surface} - E_{Pt}$, with $E_{surface-Pt}$ corresponding to Pt atoms bound to the surface, $E_{surface}$, the energy for the radical surface before Pt binding, and $E_{Pt}$, the energy of Pt atoms in vacuum. Among the studied surfaces, we find that the strongest binding for Pt atoms occurs when the substrate is the zero-layer on the SiC surface and weakest on a free-standing graphene. Using clusters composed of up to 32 atoms of Pt, the adhesion strengths per Pt atom on the zero-layer is with $\Delta E_{adsorbtion}^{32Pt-ZL/SiC} = -3.1$ eV roughly an order of magnitude larger compared to that of Pt to free-standing graphene ($\Delta E_{adsorbtion}^{32Pt-G} = 0 - .2$ eV) (see Supplementary Figure 6 and Supplementary Table 1 for DFT results). The resulting stable conformation for the large Pt cluster on the carbon zero-layer is shown in Fig. 4a. Overall, the reason of the enhanced affinity of the Pt atoms to the zero-layer is the presence of carbon atoms bound to the SiC surface which have lost their sp$^2$ character and are thereby very reactive, making the wetting of zero-layer by Pt a thermodynamically favourable process.

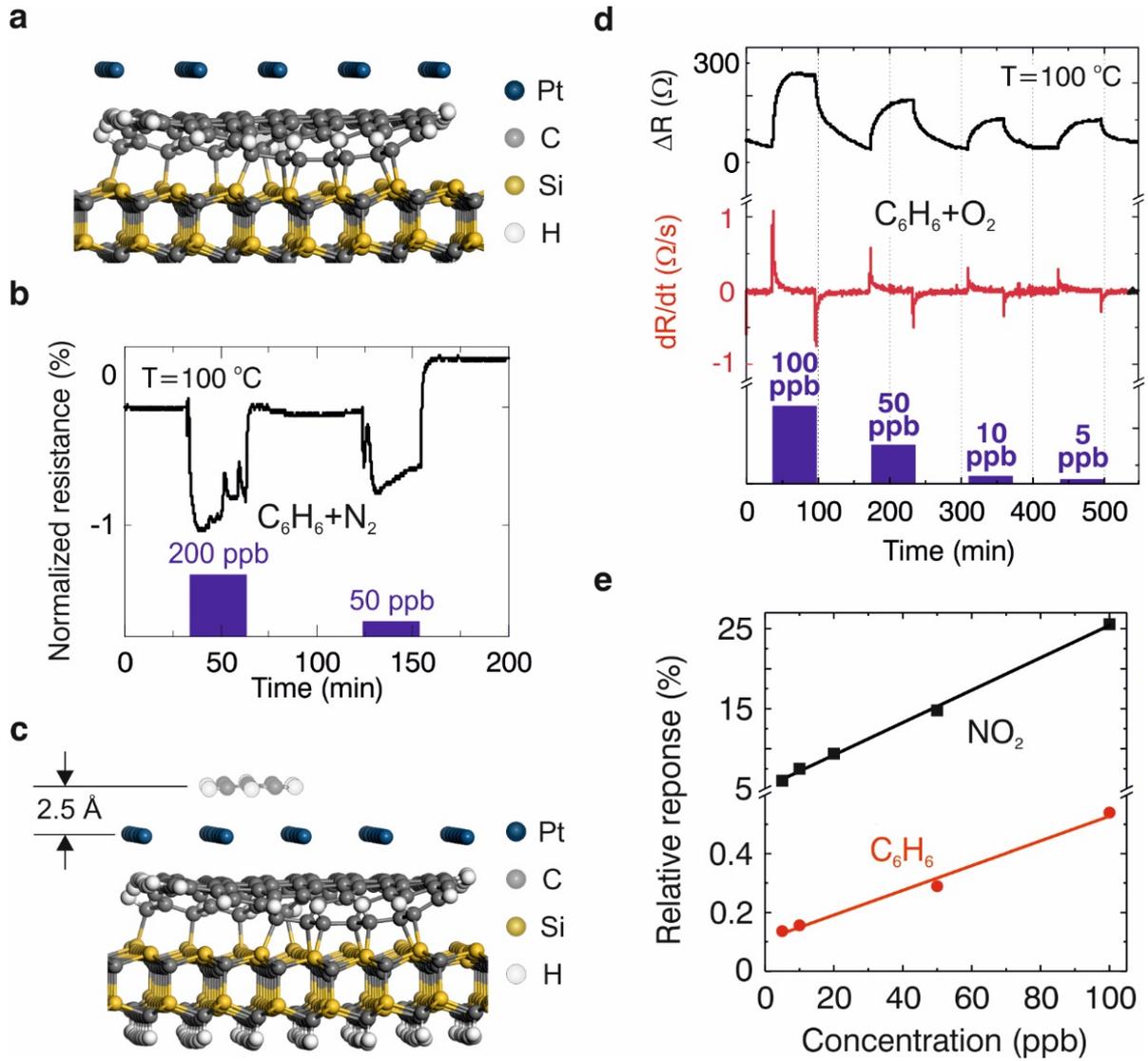

**Figure 4: Chemical to electrical transduction with atomically-thin Pt. a,** The interfacial slab is the relaxed geometry obtained from DFT calculations of a Pt monolayer on the zero-layer, partially binding to a SiC surface. For calculations, the outer carbon atoms in the zero-layer are H-terminated. In our calculations, all surfaces were geometrically relaxed before the adhesion of Pt took place **b,** Resistive read-out signal from atomically-thin Pt chemiresistor in response to benzene in $N_2$ background shown as the normalized resistance, $\hat{R} = \Delta R/R_0 = (R - R_0)/R_0$, with R the device resistance during gas exposure and $R_0 \approx 20\ \Omega$ the baseline resistance before exposure. For these measurements, the temperature is $T = 100$ °C **c,** DFT calculations help to explain the benzene detection mechanism shown in (b). This image is an interfacial slab of a geometrically-relaxed benzene molecule physisorbed on the Pt surface. The most stable conformation of benzene on Pt is that of benzene lying flat on the surface. With an equilibrium benzene – Pt distance of 2.5 Å, the adhesion strength of 0.10 eV is in the range of Van der Waals interactions and we find a hole transfer of $0.14|e|$ from the benzene molecule to Pt. **d,** Response to benzene concentrations in the range from 5 to 100 ppb at $T = 100$ °C of a large area device (5 mm x 5 mm). The black line shows the changes in device resistance upon exposure to benzene, $\Delta R = (R - R_0)$ with $R_0 \approx 39$ k$\Omega$, and the red line is its first order time-derivative. **d,** Comparison of the concentration-dependent relative response $\Delta R_{sat}/R_0 = (R_{sat} - R_0)/R_0$, with $R_{sat}$ the saturated resistance after gas exposure, for $NO_2$ ($T = 200$ °C) and $C_6H_6$ ($T = 100$ °C).

Having discussed the formation and electrical properties of atomically-thin Pt on the zero-layer, we now turn to describe the charge transfer between chemical species and Pt as the gas sensing mechanism. For this, we have fabricated chemiresistors with atomically-thin Pt (see Supplementary Figure 7) and focused on the resistive read-out of benzene, for which the interaction with Pt is well known[27,28] in order to have a firm base for comparison of the results. Fig. 4b shows the normalized resistance $\hat{R} = \Delta R/R_0 = (R - R_0)/R_0$ of a device exposed to benzene diluted in nitrogen, with R the device resistance during gas exposure and $R_0$ is the baseline resistance before exposure. The decrease of the electrical resistance of Pt chemiresistors upon exposure to benzene, shown in Fig. 4a, can be explained through the classical conductivity of the metal. Justified by our (magneto) transport measurements (Fig. 2b,c), we consider the single-band approximation of the Drude conductivity of platinum $\sigma_{Pt} = |e|p\mu$, with $e$ the elementary charge, $p$ the hole carrier density and $\mu$ the carrier mobility. Assuming that benzene adsorbed on the Pt surface does not significantly modify the mobility $\mu$ of carriers in the metallic network[29], fluctuations in the conductivity of Pt would be the consequence of changes in carrier density. Our DFT calculations point in this direction (for details see Supplementary Figure 8). Figure 4c shows geometrically optimized benzene on Pt, stabilized by Van der Waals interactions, which yield a net hole transfer of $0.14|e|$ from the benzene molecule to Pt. This corresponds to a charge transfer density of about $\Delta p_{C6H6} \approx 5 \times 10^{13}$ holes per cm$^2$, which is comparable in magnitude and carrier type to the measured baseline carrier density of Pt, $p_0$=1-3x10$^{14}$ cm$^{-2}$ (Fig. 2c). This increase in hole carrier density, $p_0 + \Delta p_{C6H6}$, allows us to qualitatively explain the increase of resistance in Pt when it interacts with benzene.

Our measurements on chemiresistor devices exposed to benzene also shed light upon the detection limits using atomically-thin Pt as sensing layers in quasi-realistic operational conditions, i.e. in a background of artificial air. Figure 4d shows the concentration-dependent response from 100 ppb down to 5 ppb, the lowest concentration attainable in our setup. Note that measurements in dry air invert the sensor response compared to measurements performed in pure N$_2$ (Fig. 4a). This resistance increase upon benzene/air exposure can be the result of oxygen/benzene competition over adsorption sites and/or through reactions with adsorbed O$^-$ [30]. In these experimental conditions, the raw response time (time to reach 90 % of steady state) is $t_r \approx$ 24 min for 5 ppb. Alternatively, $t_r$ can be evaluated based on the initial resistance increment rate, dR/dt, during gas exposure[31,32]. Defining $t_r$ as the time when dR/dt reaches the maximum value during exposure, we find $\tau_r \approx$ 12 s for 5 ppb of benzene. Now, the ultimate detection limit for benzene is estimated by considering the response of ΔR =47 Ω at 5 ppb, and assuming a linear sensitivity in the range 0 – 5 ppb. With an effective noise floor in our setup of 1.9 Ω at the operating temperature (T = 100 °C), the extrapolated lower detection limit for benzene is of the order of $1.9 \Omega/(47 \Omega/5\ ppb) = 210$ part-per-trillion. Given the charge transfer sensing mechanism, the sensitivity is higher for more strongly reducing/oxidizing gases as seen in the relative response $rr = \Delta R_{sat}/R_0 = (R_{sat} - R_0)/R_0$ in Fig. 4e, with $R_{sat}$ the saturated resistance after gas exposure. The linear sensor response in the range 100 – 5 ppb yields a sensitivity of ~0.2 %/ppb to NO$_2$, roughly 50 times higher that to benzene.

Finally, we mention that in addition to platinum, atomically-thin palladium deposited on the carbon zero-layer surface (see Supplementary Figure 9) also results in an electrically conductive surface, sensitive to the chemical environment. The well-ordered zero-layer on SiC could thus be a suitable support for the deposition of metals that have similar chemistry as the Pt- and Pd- systems[26,33]. The demonstrated gas sensing performance with atomically-thin Pt proof-of-concept devices at relatively low temperature (T = 100 °C), coupled with gas discrimination techniques[34], could already allow fast, sensitive, and selective detection of extremely toxic molecules at concentrations of relevance to, for instance, air quality monitoring. We believe that atomically-thin metals as a new platform for gas detection, aided with surface science studies, could facilitate research aiming at tuning the selectivity of gas sensing devices operating at room temperature, requirements which have proven to be challenging to meet in solid-state gas detectors. Furthermore, the carbon zero-layer, which enables the possibility to template atomically-thin metals, may also serve as a platform for epitaxy of other 2D materials and the design of new heterogeneous catalyst with superior activity and selectivity for challenging reactions[35].

**Materials & Correspondence:** should be addressed to samuel.lara@chalmers.se

**Acknowledgments**: We thank M. Skoglundh for instructive discussions, and A. Yurgens, J. F. Schneiderman, and A. Danilov for critical reading of the manuscript. S.L.A thanks Mats Hagberg for assistance with in-situ resistance measurements during Pt deposition. This work was jointly supported by the Swedish Foundation for Strategic Research (SSF) (No. IS14-0053, GMT14-0077, RMA15-0024), Knut and Alice Wallenberg Foundation, Chalmers Area of Advance NANO, the Swedish Research Council (VR) 2015-03758, the Swedish-Korean Basic Research Cooperative Program of the NRF (No. NRF-2017R1A2A1A18070721).

**Author contributions:** K.H.K, H.H. and S.L.A., contributed to device fabrication and electrical characterization. R.Y. contributed to material growth. K.L. developed the theoretical calculations. A.Z. performed surface science characterization. J.E. and M.R. conceived and performed the gas sensor tests and contributed to the design of the sensor devices. D.S. and M.P. performed STM analysis. S.L.A. and S.K conceived and designed the experiment. S.L.A and J.E. wrote the manuscript with contributions from all co-authors. All authors contributed to data evaluation and interpretation.

## Methods

### Growth

The carbon zero-layer is an integral part of the epitaxial graphene-SiC material system and is the first to form when SiC substrate is exposed to high temperature (> 1500°C in our graphene growth facilities). More specifically, this is the carbon rich surface reconstruction (6√3 x 6√3) characteristic of Si face SiC at elevated temperatures. To prevent the growth of graphene and grow only the carbon zero-layer, here we used 7 x 7 mm$^2$ 4H-SiC substrates and applied gradual heating in Ar atmosphere until 1700 °C was reached and that kept for 30 seconds. Then the furnace is switched off and the samples are taken out at room temperature. Prior to growth, the chamber it is pumped down to a base pressure of $P_0=1\times10^{-6}$ mbar in order to minimize oxygen contamination which is detrimental for a complete carbonization.

### STM

STM characterization was performed in an Aahrus STM operated at RT and at a base pressure of $1\times10^{-10}$ mbar. The tip is an etched W wire attached to the resonator of a Kolibri$^{TM}$ sensor, and we work under sample bias convention.

### Microfabrication

On the carbo zero-layer/SiC substrates, thin layers Pt atoms of thickness 3 Å to 8 Å are deposited by electron beam evaporation at a base pressure < $5\times10^{-7}$ mbar and deposition rate 1 Å/s. The zero-layers are mounted on a homemade metal holder and baked at 200 °C for 10 min right before Pt deposition to keep the surface of the zero-layer at 200 °C during evaporation. Patterning of the conductive layers into diverse geometries inclding Hall bars (30 μm x 180 μm), square bridges, and interdigitated fingers, is done by two step electron beam lithography. In the first lithography, Pd (100 nm) contacts are defined on top of the conductive layers and boundaries of the devices are carved out by reactive ion dry etching in NF3 gas using PMMA resists as a mask in the second lithography. Lithography-free chemiresistor sensor devices have been fabricated through deposition of Pd contacts on top of the Pt/zero-layer/SiC substrates using a shadow mask that has square electrodes with a 1 mm gap in between.

**Gas Sensing**

After processing, the chip was mounted on a ceramic heater substrate (Heraues GmbH, Germany) along with a Pt100 temperature sensor using ceramic glue (Aremco Ceramabond 571). A single channel source measure unit (SMU, Keithley 2601 SourceMeter) was used to drive the sensor. The inter electrode resistance was recorded at 1 Hz.

Gas exposure was controlled by a gas mixer system consisting of Bronkhorst mass flow controllers connected to PC controlled sequencing software. Flow over the sensor was constant at 100 ml/min and kept at room temperature. Humidity was added by routing part of the $N_2$ through a humidifier. The gas was kept at room temperature during all experiments. A dry background mixture of $N_2$ and $O_2$ with a ratio of 80:20 ml/min and a constant flow rate of 100 ml/min was used both as a carrier gas and purging gas. The nitrogen concentration was then adjusted when introducing test gases ($NO_2$, $C_6H_6$, $CH_2O$) to the gas flow. The sensor was exposed to test gas for 40 minutes towards 100, 50, 10 and 5 ppb, respectively, with 80 minutes of purging in a mixture of 20 % dry oxygen in a background of nitrogen

**DFT**

The energetic stability and geometrical structure of a SiC/zero-layer/Pt layer has been investigated theoretically by performing DFT calculations under periodic boundary conditions. In addition, the reactivity of this layer for benzene molecules have also been studied. More specifically, an ultrasoft pseudopotential plane-wave approach was used, based on the PBE generalized gradient approximation (GGA) of the exchange-correlation functional [1]. The GGA takes into account the gradient of the electron density, which gives a high-quality energy evaluation. To include Van der Waals interactions in the calculations, dispersion corrections were med for all calculations in the present study. Moreover, the value of the energy cutoff for the plane wave basis sets was set to 420.00 eV. In addition, the Monkhorst-Pack scheme was used for the k-point sampling of the Brillouin zone, which generated a uniform mesh of k points in reciprocal space [2]. All atoms within the interface region were allowed to move freely in the calculations by using a BFGS approach (Broyden-FletcherGoldfarb-Sharmo) [3]. The calculations in the present work were carried out using the Cambridge Sequential Total Energy Package (CASTEP) program from BIOVIA, Inc [4].

The 4H-SiC (0001) surface was used in modelling the interfacial slab. The bottom layer atoms were terminated with H in order to saturate the dangling bonds and to simulate the continuation into the bulk. All atoms, except the two bottom atomic layers, were allowed to freely relax in a geometry optimization procedure. A graphene monolayer was thereafter added to the relaxed SiC surface. In order to reduce interfacial strengths, the graphene monolayer was constructed as a very large flake, where the edge carbons were saturated with hydrogen atoms. The resulting interfacial slab was again allowed to relax, which resulted in a carbon zero-layer forming a buckled structure on the SiC surface. This theoretically obtained zero-layer was covalently binding to the underlying Si surface atoms to 20 %, which is very similar to the experimentally obtained zero-layer (15 %). This final SiC/zero-layer slab was from here on used for the growth of a monolayer 2D Pt layer, and for the reaction between gaseous benzene and this Pt layer. The calculated adsorption strengths for a 9 and 32 Pt atoms cluster yield similar results. The reason to use a large flake and not a continuous 2D Pt sheet was the presence of large strains obtained for the latter situation.

For benzene, the benzene molecule was initially positioned i) parallel, ii) orthogonal, or iii) with an angle of 45 degrees with respect to the surface. We note that for chemisorbed benzene, additionally, there is only one plausible $sp^2$-to-$sp^3$ transformation of a benzene C atom upon binding to Pt. This later however, is unstable, and as a result of geometry optimization becomes desorbed from the surface, returning to the physisorbed conformation where the benzene molecule lies flat on the Pt/zero-layer surfaces as in Fig. 4c.

**References for Methods**

# Supporting information for "Chemical sensing with atomically-thin metals templated by a two-dimensional insulator"

Kyung Ho Kim[1], Hans He[1], Marius Rodner[2], Rositsa Yakimova[2], Karin Larsson[3], Marten Piantek[4], David Serrate[5,6], Alexei Zakharov[7], Sergey Kubatkin[1], Jens Eriksson[2], Samuel Lara-Avila[1,8]*

[1] **Department of Microtechnology and Nanoscience, Chalmers University of Technology, SE-412 96, Gothenburg, Sweden**

[2] **Department of Physics, Chemistry and Biology, Linkoping University, S-581 83, Linköping, Sweden.**

[3] **Department Chemistry-Angstrom Laboratory, University of Uppsala, Uppsala, Sweden**

[4] **Instituto de Nanociencia de Aragón and Laboratorio de Microscopías Avanzadas, Universidad de Zaragoza, 50018 Zaragoza, Spain.**

[5] **Instituto de Ciencia de Materiales de Aragón, CSIC-University of Zaragoza, 50009, Zaragoza, Spain**

[6] **Departamento de Física de la Materia Condensada, University of Zaragoza, 50009, Zaragoza, Spain**

[7] **MAX IV Laboratory and Lund University, 221 00 Lund, Sweden.**

[8] **National Physical Laboratory, Hampton Road, Teddington, TW11 0LW, UK**

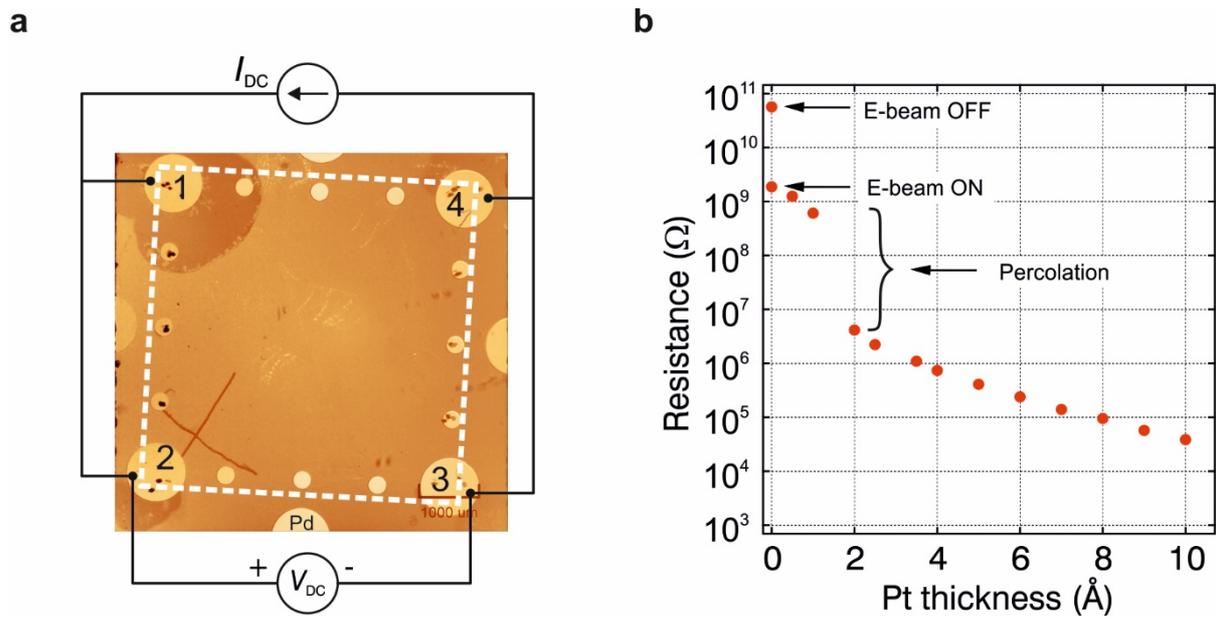

**Supplementary Figure 1. In-situ measurements of Pt resistance vs Pt thickness on a 7 mm x 7mm carbon zero-layer/SiC substrate. a,** Schematic of the electrical connections and optical micrograph of the 7 x 7 mm² chip used for in-situ resistance measurements; for measurements, voltage bias is applied the resistance is measured between pre-deposited Pd contacts 1-2 and 3-4. Pt is deposited through a shadow mask with geometry indicated by the dashed white square. Measurements were performed in a Lesker PVD 225 electron beam evaporation system, fitted with electrical feedthroughs for two-terminal resistance measurements during Pt deposition. **b,** The initial resistance (zero-bias differential resistance) of the substrate is of the order of ~50 GΩ which drops to few GΩ after switching on the electron gun (SiC is photoconductive, $E_G = 3.21$ eV). The two-terminal resistance decreases slowly for the first 1.5 Å, followed by a sudden drop of two orders of magnitude (to a few MΩ) at thickness $t_{Pt} = 2$ Å. The deposition condition are $P = 1\times10^{-7}$ mbar, $T_{substrate} = 300$K, and deposition rate r = 0.017 Å/s.

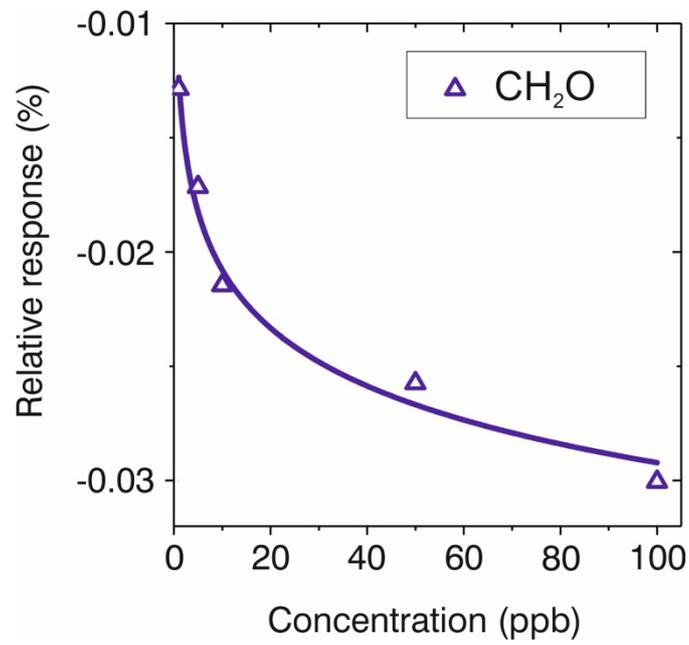

**Supplementary Figure 2. Atomically-thin Pt chemiresistor response to formaldehyde (CH₂O).** The relative response $\hat{R} = \Delta R/R_0 = (R - R_0)/R_0$ to CH$_2$O is found to be non-linear in the range 5 to 100 ppb, likely due to Langmuir adsorbtion[R1].

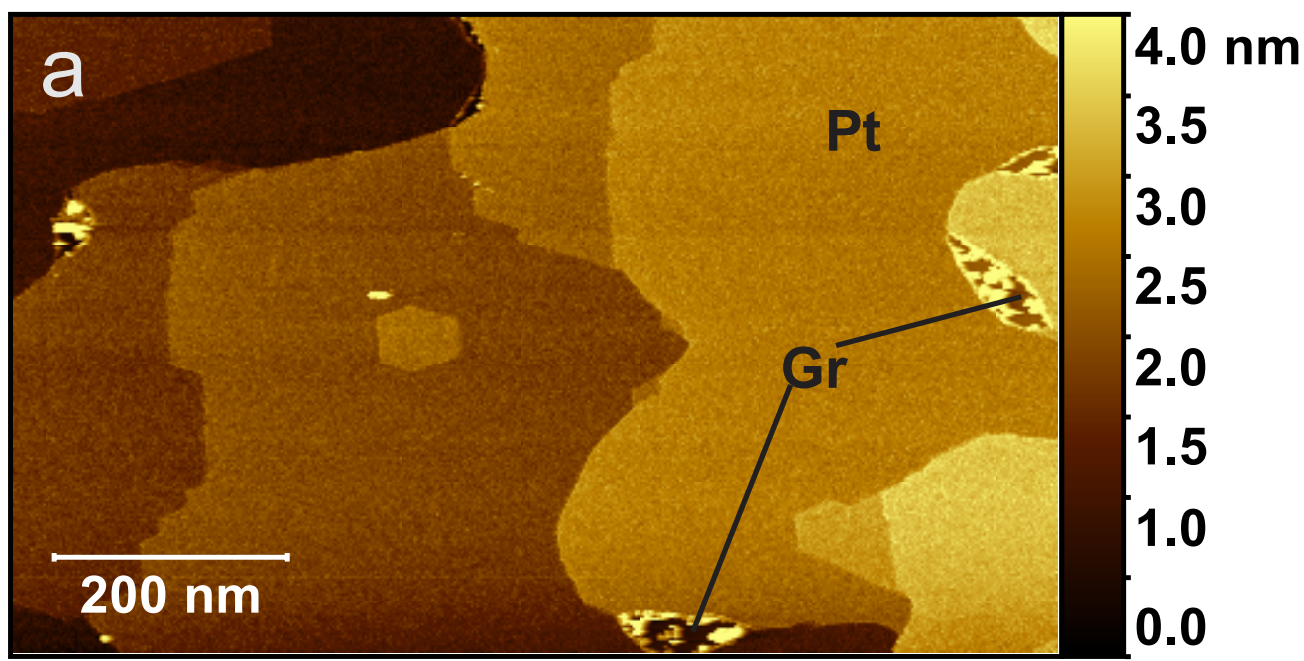

**Supplementary Figure 3**. **Scanning tunneling microscopy (STM) on 3Å Pt deposited on the carbon zero-layer on SiC.** Constant current STM topographies (regulation 1 V and 60 pA) showing an overview of the Pt covered surface. In addition to the reconstructed steps on the SiC surface, it is possible to find small graphene inclusions (Gr) that, unlike the carbon zero-layer, remain atomically clean.

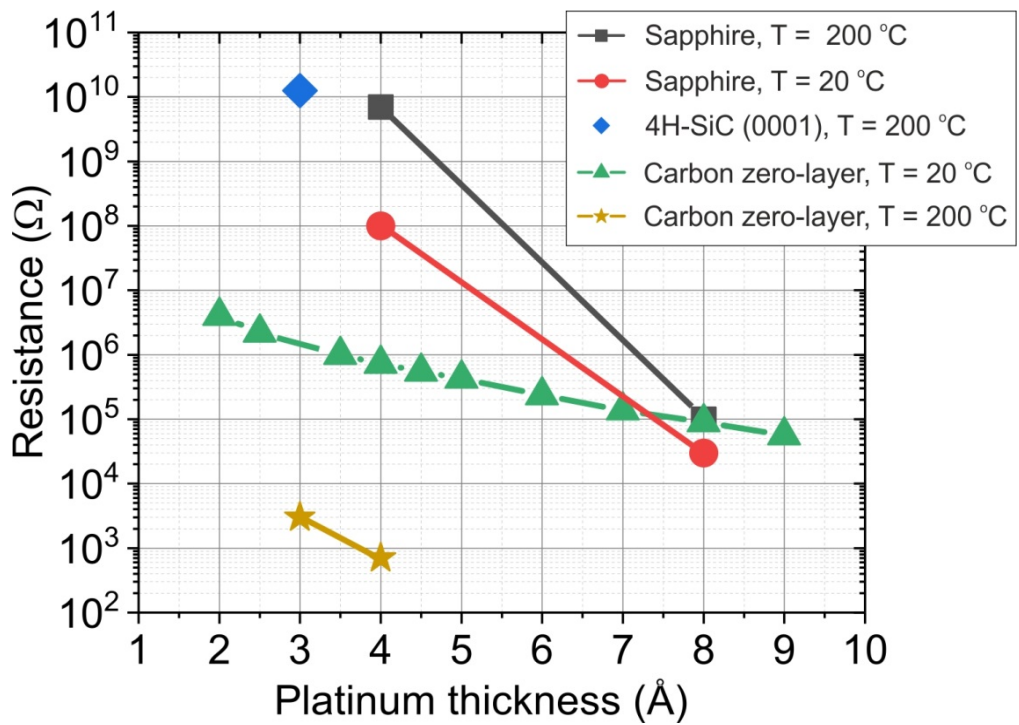

**Supplementary Figure 4**. **Deposition of Pt on different substrates.** Comparison of the resistance of Pt deposited on the C plane of sapphire substrates, bare SiC, and the carbon zero-layer on SiC, at temperature 200 °C and 20 °C. The resistance is measured over the entire substrate measured as indicated in Supplementary figure 1. The sizes of the substrates are 7 x 7 mm$^2$ (SiC) and 6 x 6 mm$^2$ (sapphire).

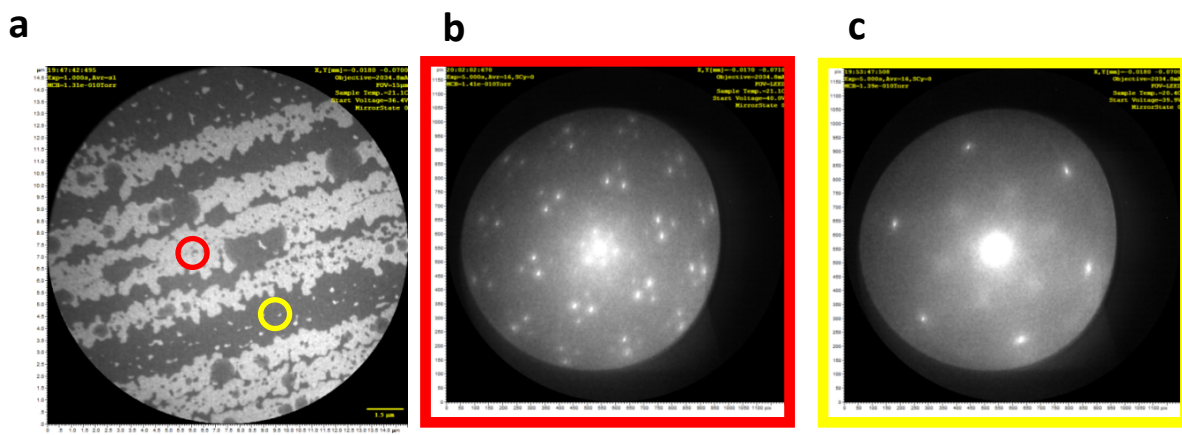

**Supplementary Figure 5. Low energy electron microscopy (LEEM) on incomplete carbon zero-layer covered with 3Å Pt (Field of View=15 μm, E=36.4eV). a,** The contrast is due to different phases on the surface: white areas correspond to the carbon-rich zero-layer (6√3 x 6√3 phase) while dark areas is a silicon-rich SiC. The contrast in LEEM is nicely seen through 3Å of Pt. **b,** corresponding micro-low energy electron diffraction (μ-LEED) of the Pt-covered carbon rich zero-layer (red circle in a). **c,** μ-LEED of the Pt-covered silicon-rich SiC substrate (yellow circle in (a)). In both cases, the attenuation of the LEED spots intensity confirms the thickness (~3Å) of the Pt film.

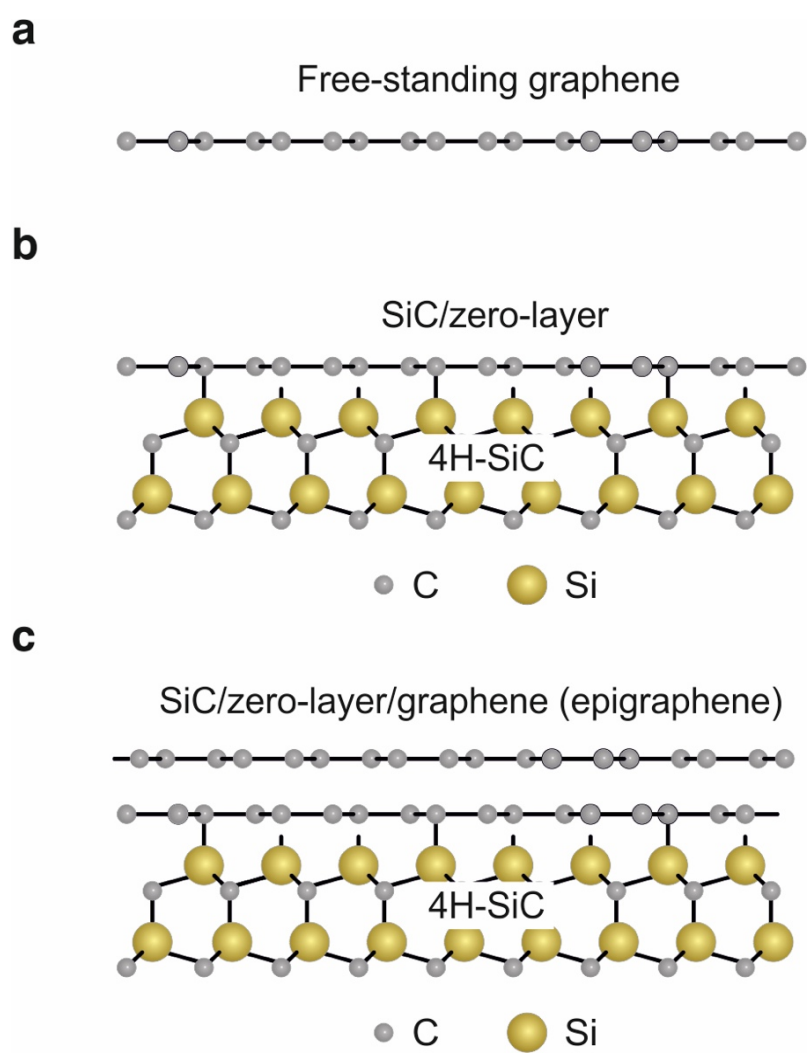

**Supplementary Figure 6. Substrates used in DFT calculations to investigate the growth of Pt. a,** Free-standing graphene. **b,** The $6\sqrt{3} \times 6\sqrt{3}$ carbon-rich zero-layer on SiC. **c,** Epigraphene, composed of SiC substrate, carbon zero-layer and a graphene layer as the topmost surface.

|            | ΔE$_{adhesion}$ (eV) | | |
|---|---|---|---|
| # Pt atoms | Free-standing graphene | SiC/zero-layer | SiC/zero-layer/graphene |
| 1Pt        | -5.6         | -9.6          | -9.3          |
| 9 Pt flake | -2.2 (-0.2)* | -27.0 (-3.0)* | -2.9 (-0.3)*  |
| 32 Pt flake| -6.4 (-0.2)* | -99.2 (-3.1)* | -9.3 (-0.3)*  |

**Supplementary Table 1. Adhesion energies for Pt on various two-dimensional carbon materials of Supplementary Figure 6 from DFT calculations.** Our results include energies calculated for adhesion of three sizes of Pt clusters onto a) a free-standing graphene sheet, b) a SiC/zero-layer surface, and c) a SiC/zero-layer/graphene (epigraphene) surface. The three Pt clusters include 1, 9, and 32 Pt atoms, respectively. The adhesion energies are defined as ΔE$_{adhesion}$ = E$_{surface-Pt}$ − E$_{surface}$ - E$_{Pt}$ were E$_{surface-Pt}$, E$_{surface}$ and E$_{Pt}$ are the optimised energy for i) a Pt binding to the surface, ii) the radical surface before Pt binding, and iii) a single Pt atom, respectively. These total energies were based on a model that was geometry optimized Indicated in starred parenthesis are the adhesion energies per Pt atom. A single Pt atom was found to bind with two bonds to both SiC/zero-layer and SiC/zero-layer/graphene. As a comparison, the chemisorption to an on-top position on a free-standing graphene layer was much weaker (5.6 eV). For larger number of Pt atoms, the Pt-to-surface bond is now a compromise between the intra-flake Pt-Pt bonds and the Pt-surface bonds (i.e., Pt-Si bonds). As can be seen in the Table, the calculated adhesion strengths per atom are almost identical for 9 and 32 Pt atoms in all cases. The reason to use a large flake and not a continuous 2D Pt sheet was the presence of large strains obtained for the latter situation

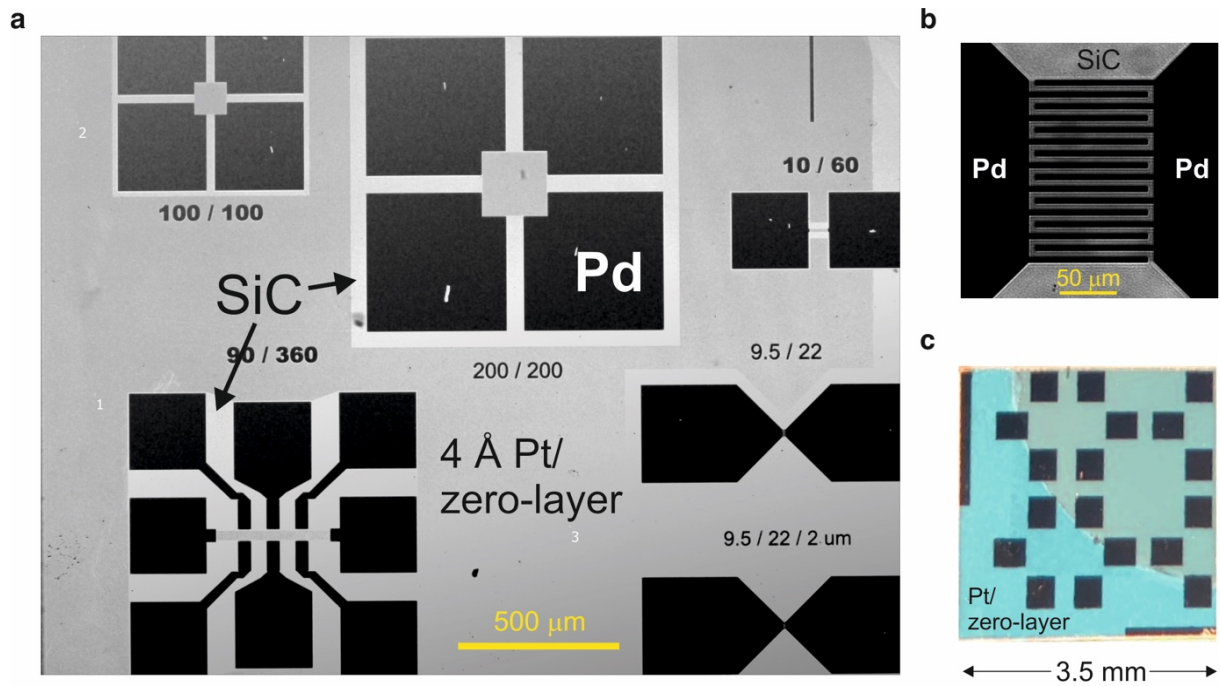

**Supplementary Figure 7. Microfabrication of devices. a,** Optical micrograph in transmission mode showing examples of devices made with atomically-thin Pt. Pd contacts (black) are deposited after the first electron beam (e-beam) lithography layer, and the second and last e-beam lithography step is to remove Pt (bright contrast) by reactive ion etching, to carve out the desired device geometry. **b,** A zoomed-in picture of a interdigitated device, with 100 μm-long , 10 μm-wide fingers. **c,** large devices made in which the Pd contacts are deposited through a shadow mask to avoid e-beam lithography.

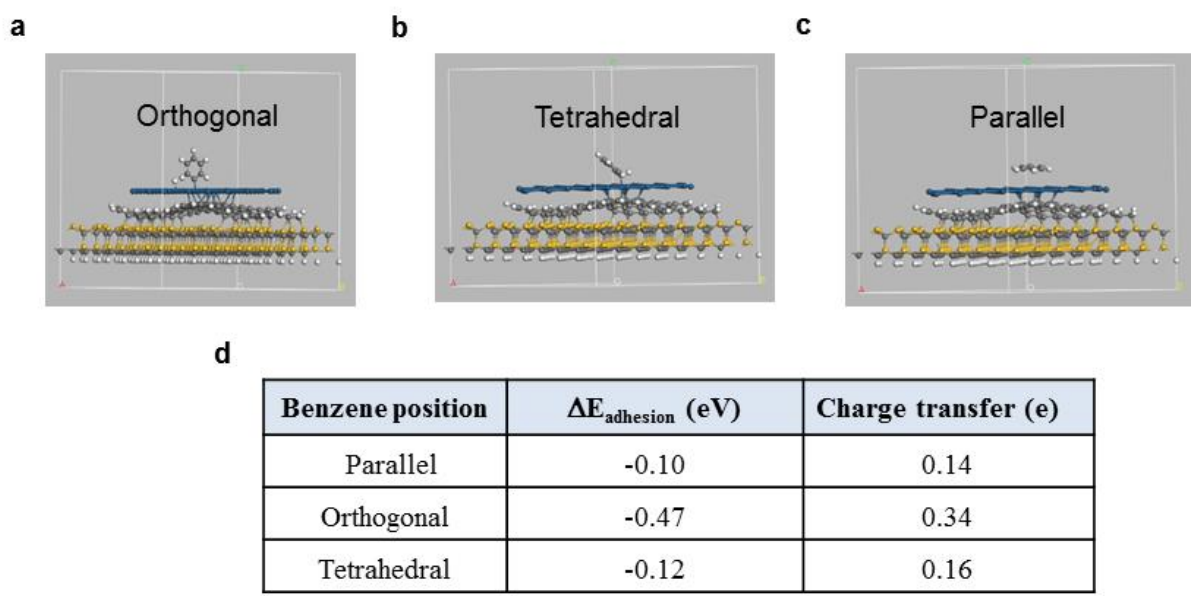

| Benzene position | $\Delta E_{adhesion}$ (eV) | Charge transfer (e) |
|---|---|---|
| Parallel | -0.10 | 0.14 |
| Orthogonal | -0.47 | 0.34 |
| Tetrahedral | -0.12 | 0.16 |

**Supplementary Figure 8. Summary of DFT calculations for benzene adsorbtion on Pt/zero-layer.** Interfacial slabs showing a benzene molecule chemisorbed to a SiC/zero-layer/Pt surface: **a,** Final geometry of benzyl and an H radical chemisorbed to neighbouring surface sites (with benzyl in an orthogonal position); **b,** Initial geometry of a chemisorbed benzene molecule were the binding benzene carbon has become $sp^3$-hybridized (with benzene forming a tetrahedral angle to the surface); **c,** Final geometry of the model shown in (b), after geometry optimization calculations (as a result of the geometry optimization, all the initial benzene positions transferred to a parallel position above the SiC/zero-layer/Pt surface). The color coding is: Pt = blue; C = grey; Si = orange; H = white. **d,** Adhesion energies for the attachment of benzene to a SiC/zero-layer/Pt surface in different geometries including orthogonal chemisorption of benzyl and H (**a**), tetrahedral chemisorption of benzene with an sp3 –hybridized C atom (binding to the surface) (**b**), and parallel position of native benzene (**c**).

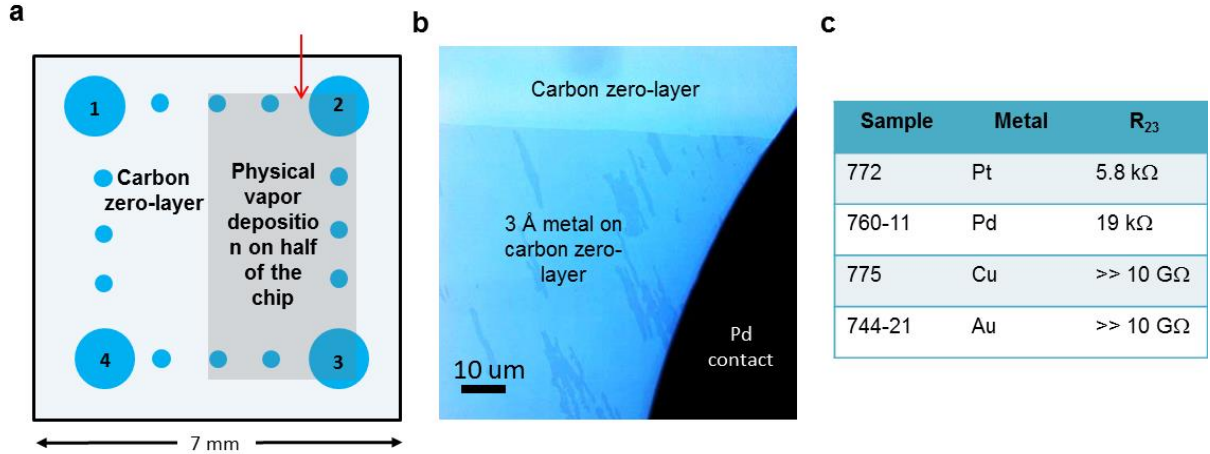

**Supplementary Figure 9. Deposition of other metals on the carbon zero-layer. a,** For all tested metals, deposition is made on only half of the chip (7 mm x 7 mm) for later comparison of the metal-free and metallized zero-layer surfaces. Contacts for electrical measurements (circles) can be deposited before or after the thin metal layers, we have observed no difference in contact performance. **b,** the deposition of metals can be verified with optical microscope in transmission mode[R2]. The Pd contact appears black (no light transmission), and the brighter contrast correspond to metal-free carbon zero-layer. **c,** summary of representative resistances obtained when depositing 4Å of different metals on the carbon zero-layer surface. The quoted resistance values correspond to the two-terminal resistance measured between contacts 2 and 3 in (a). With the given deposition conditions, R = 1 Å/s and T = 200 C, Pt and Pd form an electrically conductive layer over the entire substrate.

**Supporting references**